\documentclass[twocolumn,epjc3]{svjour3}          % twocolumn

\RequirePackage[T1]{fontenc}

\smartqed  % flush right qed marks, e.g. at end of proof

\usepackage{amssymb}
\RequirePackage{graphicx}
\RequirePackage{mathptmx}      % use Times fonts if available on your TeX system
\RequirePackage{flushend}
\RequirePackage[numbers,sort&compress]{natbib}
\RequirePackage[colorlinks,citecolor=blue,urlcolor=blue,linkcolor=blue]{hyperref}

\journalname{Eur. Phys. J. C}

\begin{document}

\title{A polarisation modulation scheme for measuring vacuum magnetic birefringence with static fields}

\author{G. Zavattini\thanksref{e1,addr1}
        \and
        F. Della Valle\thanksref{addr2}
        \and
        A. Ejlli\thanksref{addr1}
        \and
        G. Ruoso\thanksref{addr3} %etc.
}

\thankstext{e1}{e-mail: guido.zavattini@unife.it}

\institute{INFN, Sez. di Ferrara and Dip. di Fisica e Scienze della Terra, Universit\`a di Ferrara, via G. Saragat 1, Edificio C, I-44122 Ferrara (FE), Italy\label{addr1}
          \and
          INFN, Sez. di Trieste and Dip. di Fisica, Universit\`a di Trieste, via A. Valerio 2, I-34127 Trieste (TS), Italy\label{addr2}
          \and
          INFN, Lab. Naz. di Legnaro, viale dell'Universit\`a 2, I-35020 Legnaro (PD), Italy\label{addr3}
}

\date{Received: date / Accepted: date}

\maketitle

\begin{abstract}
A novel polarisation modulation scheme for polarimeters based on Fabry-Perot cavities is presented. The application to the proposed HERA-X experiment aiming to measuring the magnetic birefringence of vacuum with the HERA superconducting magnets is discussed. 
\end{abstract}

%\pacs{07.60.Fs, 12.20.-m, 42.50.Xa, 78.20.Ls}
%\noindent{\it magneto-optical tests of QED; nonlinear electrodynamics; PVLAS\/}
%\maketitle

\section{Introduction}

Vacuum magnetic birefringence is a non linear electrodynamic effect in vacuum closely related to light-by-light elastic scattering. Predicted by the Euler-Heisenberg-Weisskopf effective Lagrangian density \cite{EHW,Delta_n_QED} written in 1936,

\begin{eqnarray*}
{\cal L}_{\rm EHW}&=&\frac{1}{2\mu_{\rm 0}}\left(\frac{E^{2}}{c^{2}}-B^{2}\right)+\\
&+&\frac{A_{e}}{\mu_{\rm 0}}\left[\left(\frac{E^{2}}{c^{2}}-B^{2}\right)^{2}+7\left(\frac{\vec{E}}{c}\cdot\vec{B}\right)^{2}\right],
\end{eqnarray*}
%{\footnotesize\[
%{\cal L}_{\rm EHW} = \frac{1}{2\mu_{\rm 0}}\left(\frac{E^{2}}{c^{2}}-B^{2}\right)+\frac{A_{e}}{\mu_{\rm 0}}\left[\left(\frac{E^{2}}{c^{2}}-B^{2}\right)^{2}+7\left(\frac{\vec{E}}{c}\cdot\vec{B}\right)^{2}\right]
%\]}
it takes into account the vacuum fluctuations of electron-positron pairs. As of today, ${\cal L}_{\rm EHW}$ still needs experimental confirmation. Here
 \[
A_e=\frac{2}{45\mu_{0}}\frac{\alpha^2 \mathchar'26\mkern-10mu\lambda_e^{3}}{m_{e}c^{2}}=1.32\times 10^{-24} {\rm{~T}}^{-2}.
\]
The ellipticity $\psi$ induced on a linearly polarised beam of light with wavelength $\lambda$ passing through a medium with birefringence $\Delta n$ and length $L$, and whose axes are defined by the external magnetic field, is
\[
\psi=\pi\frac{L}{\lambda}\Delta n \sin2\phi
\]
where $\phi$ is the angle between the magnetic field and the polarisation direction. The birefringence predicted by ${\cal L}_{\rm EHW}$ is \cite{Delta_n_QED}
\[
\Delta n=3A_e B^2 \simeq 4\times10^{-24}B^2.
\]

Several experiments are underway, of which the most sensitive at present are based on polarimeters with very high finesse Fabry-Perot cavities and variable magnetic fields \cite{DellaValle2015,Q&A2010,BMV2014}. The Fabry-Perot cavity is necessary to increase the optical path $L$ within the magnetic field region, whereas the variable magnetic field is necessary to induce a time dependent effect. Both of these aspects significantly increase the sensitivity of the polarimeters.% Two schemes have been proposed to vary the magnetic field: rotating permanent magnets with a field of $B = 2.5$~T and high field pulsed magnets with a magnetic field $B\simeq15$~T lasting a couple of milliseconds.

High field static superconducting magnets such as those used in the LHC and HERA accelerators have also been proposed but their use is limited by the difficulty in modulating, in one way or another, their magnetic fields. To work around this problem, proposals of rotating the polarisation have been considered \cite{OSQAR2005}, but the presence of the Fabry-Perot cavity, whose mirrors always present an intrinsic birefringence whose induced ellipticity is orders of magnitude larger than the ellipticity due to vacuum magnetic birefringence, have made this idea unfeasible.

In this note, a novel modulation scheme is presented that might profitably be employed with large superconding magnets.

\section{Preliminary considerations}

In a recent workshop in Hamburg \cite{QED2015}, a new scheme, presented in this paper, has been suggested to measure the magnetic birefringence of vacuum predicted on the basis of the 1936 effective Lagrangian ${\cal L}_{\rm EHW}$.
The HERA-X experiment \cite{HERA-X} proposes to make use of the powerful infrastructure of the \mbox{ALPSIIc} set-up \cite{ALPS2c}: about 5000~T$^2$m, which could go up to about 7700~T$^2$m if the peak field of 6.6~T is employed. The magnetic birefringence in HERA-X will therefore be
\[
\Delta n^{\mbox{\scriptsize(HERA-X)}}\approx10^{-22}
\]
for the 5.3~T magnetic field. With this birefringence, the maximum ellipticity
\[
\psi^{\mbox{\scriptsize(HERA-X)}}=\pi\frac{L}{\lambda}\Delta n^{\mbox{\scriptsize(HERA-X)}}
\]
is $\psi^{\mbox{\scriptsize(HERA-X)}} = 5\times 10^{-14}$ for $\lambda=1064$~nm and $L = 177$~m. In the usual setups, the magnetic field is modulated to gain sensitivity. In the particular case of the HERA superconducting magnets the electric current in can be modulated at about a millihertz frequency \cite{Trines2015}.
\begin{figure}[htb]
\begin{center}
\includegraphics[width=8.5cm]{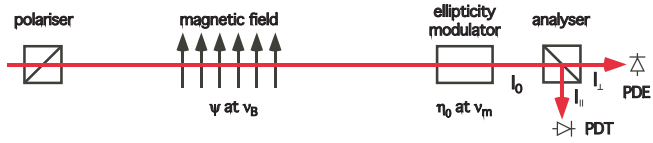}
\end{center}
\caption{A simple heterodyne ellipsometer. PDE: Extinction Photodiode; PDT: Transmission Photodiode.}
\label{heterodyne}
\end{figure}

Let's analyse the measurement scheme of figure \ref{heterodyne}, featuring two crossed polarisers, a variable magnetic field (fixed direction) at a frequency $\nu_{\rm B}$, and an ellipticity modulator at a frequency $\nu_{\rm m}$ for linearising the effect. In this scheme the intensity collected at the photodiode PDE is, at the lowest useful order,
\[
I_\perp(t)\simeq I_0\,\left[\eta^2(t)+2\eta(t)\psi(t)\right] + \mathcal{O}\left[\psi(t)^2\right].
\]
The interesting signal is found, in a Fourier transform of the signal from the photodiode, at the two sidebands $\pm\nu_{\rm B}$ from the carrier frequency $\nu_{\rm m}$ of the ellipticity modulator. 

The resulting peak shot-noise sensitivity in such a scheme is
\[
S_{\rm shot}=\sqrt{\frac{2e}{I_{0}q}},
\]
where $e$ is the electron charge, $I_0$ is the intensity reaching the analyser, and $q$ is the quantum efficiency of the photodiode. With $I_0\simeq100$~mW and $q=0.7$~A/W, the shot-noise peak sensitivity is $S_{\rm shot} \simeq 2\times 10^{-9}~1/\sqrt{\rm Hz}$.
Despite the exceptional parameters of the magnetic field of HERA-X, the integration time $T$ to achieve a unitary signal-to-noise ratio remains too long, even supposing to work at shot-noise sensitivity:
\[
T\sim\left(\frac{S_{\rm shot}}{\psi^{\mbox{\scriptsize(HERA-X)}}}\right)^2\sim10^{9}~{\rm s}.
\]
As mentioned above, further amplification is required. This can be achieved with a Fabry-Perot cavity, which can be thought of as a lengthening of the optical path by a factor $N=2{\cal F}/\pi$, where ${\cal F}$ is the finesse of the cavity. The proposed finesse for HERA-X is ${\cal F}=60,000$. With such a finesse, the ellipticity $\psi$ increases by a factor $N = 38,000$ and the integration time therefore diminishes by a factor $N^2$. Assuming shot-noise sensitivity, on paper, this device should easily allow the measurement.

\begin{figure}[htb]
\begin{center}
\includegraphics[width=8.5cm]{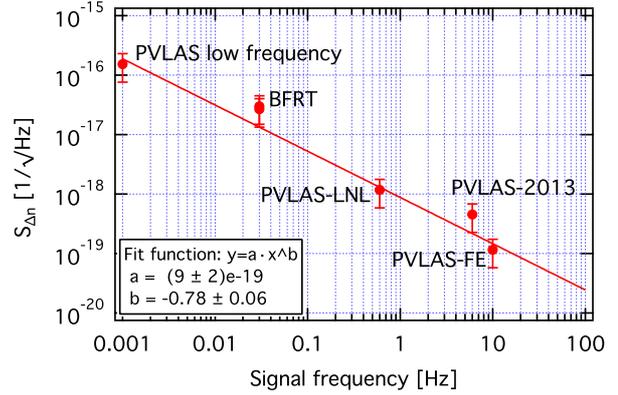}
\end{center}
\caption{Birefringence noise densities measured in polarimeters set up to measure the magnetic vacuum birefringence plotted as function of the frequency. Data from the experiments BFRT \cite{Cameron1993}, PVLAS-LNL \cite{ Bregant2008}, PVLAS-2013 \cite{DellaValle2013}, PVLAS-FE \cite{DellaValle2015} are normalised to the length of the optical cavities, to the number of passes and to the wavelength. The leftmost point has been measured during the 2015 data taking campaign of the PVLAS experiment. The two almost equivalent points from BFRT are measured with two different cavities, one having 34 passes and the other 578 passes. The error bars are an estimated 50\%.}
\label{noise}
\end{figure}

A problem remains, however, regarding the actual sensitivity that one may reasonably think to achieve at low frequencies with such a long cavity. Let us consider the experiments on this subject realised so far with a scheme similar to the one proposed with HERA-X \cite{Cameron1993,Bregant2008,DellaValle2013,DellaValle2015}. In Figure \ref{noise} we show the noise densities in birefringence
\[
S_{\Delta n} = S_{\psi}\frac{\lambda}{\pi\left(\frac{2{\cal F}}{\pi}\right)d}
\]
measured in these apparatuses as a function of the frequency of the effect. In this formula $S_{\psi}$ is the ellipticity sensitivity of each experiment, $\lambda$ is the wavelength, ${\cal F}$ is the finesse and $d$ the cavity length. Note that the cavity length $d$ has been used instead of the length $L$ of the magnetic region; what is plotted is therefore the {\em best} sensitivity in birefringence that {\em could} be obtained by the experiments. In the figure we did not report a much higher sensitivity value of the Q~$\&$~$\!$A experiment \cite{Q&A2010}. The data are fitted with a power function.

The message put forward by Figure \ref{noise} is that increasing the effective length (finesse and magnetic field length) does not guarantee the shortening of the necessary integration time to reach a unitary signal-to-noise ratio; seeking the highest finesse possible is not necessarily the optimal choice. Increasing the birefringence modulation frequency seems to be more effective. Furthermore, with lower finesses, the cavity will have a shorter decay time and therefore a higher cutoff frequency allowing higher modulation frequencies. Figure \ref{noise} suggests, therefore, that the finesse of the cavity should be the highest for which the polarimeter is still limited by intrinsic noises.

The figure suggests that it is unlikely that, at 1~mHz, a sensitivity better than $S_{\Delta n}^{\rm (1~mHz)}\approx 10^{-16}/\sqrt{\rm Hz}$ can be reached. As a matter of fact, the sensitivity of a giant 200~m cavity, necessarily built with the end mirrors sitting on separate benches, can hardly be predicted. Even assuming for HERA-X the sensitivity of Figure \ref{noise} at 1~mHz, a SNR~=~1 could only be reached in about
\[
T = \left(\frac{S_{\Delta n}^{\rm (1~mHz)}}{\Delta n^{\mbox{\scriptsize(HERA-X)}}}\right)^2 \approx 10^{12}~{\rm s}.
\]
%This is a fairly short integration time but depends entirely on the possibility to reach such sensitivities with such a long cavity. 
%One can easily convince oneself that increasing the effect ($\psi$) never guaranteed the shortening of the necessary integration time to reach a SNR~=~1. The increase of the modulation frequency of the effect ($\nu_B$) always proved to be more effective.

\section{Method}

\begin{figure}[htb]
\begin{center}
\includegraphics[width=8.5cm]{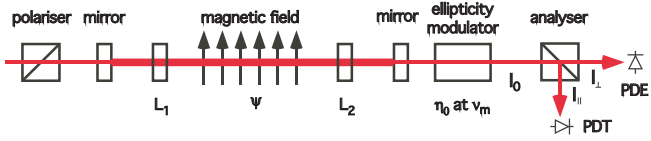}
\end{center}
\caption{Proposed modulation scheme. %M$_{1,2}$: cavity mirrors. 
L$_{1,2}$: rotating half-wave-plates. PDE: Extinction Photodiode; PDT: Transmission Photodiode.}
\label{Scheme}
\end{figure}

In this note, we present a novel modulation scheme that would bring in several advantages. This idea has never been tested in a laboratory, but is likely to be more effective than the one described above. In this way one can work at higher frequencies for the best sensitivity. In this scheme the magnetic field does not need to be modulated. The scheme consists in introducing a pair of co-rotating half-wave-plates $L_1$ and $L_2$ \emph{inside the Fabry-Perot cavity}, as schematically shown in Figure \ref{Scheme}. The polarisation within the magnetic field would rotate at twice the frequency of the wave-plates and should allow to increase substantially the modulation frequency of the effect. %In this way the polarimeter may work with a greatly reduced noise level.
An important feature of this scheme is that the polarisation direction of the light on the Fabry-Perot mirrors would remain fixed, thereby eliminating the contribution of the ellipticity due to the intrinsic birefringence of the mirrors. Furthermore, the polarisation direction on each mirror could be chosen; the input polariser defines the polarisation direction on the first mirror whereas on the second mirror the polarisation direction is defined by the relative angle between the axes of the two wave-plates.

Let us indicate with $\nu_{\rm L}$ the rotation frequency of the wave-plates, that we suppose to rotate synchronously but not necessarily aligned one to the other. The Jones representation of the electric field at the exit of the cavity is

\begin{eqnarray*}
\mathbf{E}_{\rm out}(\delta)&=&\left(\begin{array}{c}E_{{\rm out},\parallel}\\E_{{\rm out},\perp}\end{array}\right)=\\
&=&E_0\,\sum_{n=0}^\infty\left[Re^{i\delta}\,\mathbf{L}_2\cdot\mathbf{X}\cdot\mathbf{L}_1^2\cdot\mathbf{X}\cdot\mathbf{L}_2\right]^n\cdot\\
&&Te^{i\delta/2}\,\mathbf{L}_2\cdot\mathbf{X}\cdot\mathbf{L}_1\cdot\left(\begin{array}{c}1\\0\end{array}\right)=\\
&=&E_0\,\left[\mathbf{I}-Re^{i\delta}\,\mathbf{L}_2\cdot\mathbf{X}\cdot\mathbf{L}_1^2\cdot\mathbf{X}\cdot\mathbf{L}_2\right]^{-1}\cdot\\
&&Te^{i\delta/2}\,\mathbf{L}_2\cdot\mathbf{X}\cdot\mathbf{L}_1\cdot\left(\begin{array}{c}1\\0\end{array}\right),
\end{eqnarray*}
%{\footnotesize 
%\begin{eqnarray*}
%&\mathbf{E}_{\rm out}(\delta)=\left(\begin{array}{c}E_{{\rm out},\parallel}\\E_{{\rm out},\perp}\end{array}\right)=\\
%&E_0\,\sum_{n=0}^\infty\left[Re^{i\delta}\,\mathbf{L}_2\cdot\mathbf{X}\cdot\mathbf{L}_1^2\cdot\mathbf{X}\cdot\mathbf{L}_2\right]^n\cdot Te^{i\delta/2}\,\mathbf{L}_2\cdot\mathbf{X}\cdot\mathbf{L}_1\cdot\left(\begin{array}{c}1\\0\end{array}\right)\\
%&=E_0\,\left[\mathbf{I}-Re^{i\delta}\,\mathbf{L}_2\cdot\mathbf{X}\cdot\mathbf{L}_1^2\cdot\mathbf{X}\cdot\mathbf{L}_2\right]^{-1}\cdot Te^{i\delta/2}\,\mathbf{L}_2\cdot\mathbf{X}\cdot\mathbf{L}_1\cdot\left(\begin{array}{c}1\\0\end{array}\right),
%\end{eqnarray}}
where $\delta$ is the round-trip phase acquired by the light between the two cavity mirrors, $R$ and $T$ are the reflectivity and transmissivity of the mirrors, $\mathbf{I}$ is the identity matrix,
\[
\mathbf{X}=\left(\begin{array}{cc}e^{i\psi}&0\\0&e^{-i\psi}\end{array}\right)
\]
is the magnetic birefringence of vacuum generating an ellipticity $\psi$ in the polarisation of the light, and
\[
\mathbf{L}_{1,2}=\mathbf{R}(-\phi-\phi_{1,2})\cdot\mathbf{L}_0(\pi+\alpha_{1,2})\cdot\mathbf{R}(\phi+\phi_{1,2})
\]
are the rotating wave-plates. Here
\[
\mathbf{L}_0(\alpha)=\left(\begin{array}{cc}e^{i\alpha/2}&0\\0&e^{-i\alpha/2}\end{array}\right)
\]
represents the wave-plate and
\[
\mathbf{R}(\phi)=\left(\begin{array}{cc}\cos\phi&\sin\phi\\-\sin\phi&\cos\phi\end{array}\right),
\]
the rotation matrix, with $\phi$ the variable azimuthal angle of the wave-plates: $\phi(t)=\nu_{\rm L}t$. The angle $\phi_2-\phi_1$ is the constant relative phase between the slow axes of the two rotating wave-plates, and $\alpha_{1,2}$ allow for small deviations from $\pi$ of the retardation of the two imperfect wave-plates. The electric field after the analyser is then
\[
\mathbf{E}(\delta)=\mathbf{A}\cdot\mathbf{H}\cdot\mathbf{R}(2\phi_2-2\phi_1)\cdot\mathbf{E}_{\rm out}(\delta),
\]
where
\[
\mathbf{H}=\left(\begin{array}{cc}1&i\eta(t)\\i\eta(t)&1\end{array}\right)\qquad\mbox{and}\qquad
\mathbf{A}=\left(\begin{array}{cc}0&0\\0&1\end{array}\right)
\]
are the ellipticity modulator, placed at $45^\circ$ with respect to the output polarisation, and the analyser set to maximum extinction, respectively. In the expression for $\mathbf{H}$, $\eta(t)=\eta_0\cos2\pi\nu_{\rm m} t$. The rotation matrix between the cavity and the ellipticity modulator ensures that the modulator and the analyser are correctly oriented. To first order in $\alpha_1$, $\alpha_2$, and $\psi$, the intensity detected by the photodiode PDE is given by
\begin{eqnarray*}
&&I(\delta)\approx I_0\,\displaystyle\frac{T^2}{1-R\cos\delta+R^2}\times\\
&&\times\left\{\eta(t)^2+\frac{2\,\eta(t)\,(1-R)}{1-R\cos\delta+R^2}\,\Big[\right.\psi\sin(4\phi(t)+4\phi_1)+\\
&&+\left.\alpha_1\sin(2\phi(t)+2\phi_1)+\alpha_2\sin(2\phi(t)+4\phi_1-2\phi_2)\Big]\right\}.
\end{eqnarray*}

An interesting result from this formula is that the signal of the magnetic birefringence of vacuum is found at the frequencies $\nu_m\pm4\nu_{\rm L}$ deriving from the product $\eta(t)\,\psi\sin(4\phi(t)+4\phi_1)$, while the signals due to imperfect wave-plates come at $\nu_{\rm m}\pm2\nu_{\rm L}$. In the above formulas we have not considered the intrinsic birefringence of the mirrors \cite{Zavattini2006}. In this scheme, by choosing appropriately $\phi_1$ and $\phi_2$ it should be possible to minimise the effect of this birefringence by independently aligning, on each mirror, the polarisation of the light to the birefringence axes of the mirrors \cite{DellaValle2015}.

Clearly the presence of the two half-wave-plates inside the cavity introduces some losses. Therefore there is an upper limit to the finesse one can obtain due to the absorption of the wave-plates. With a correct antireflective coating, wave-plates can be obtained with a total absorption of $\simeq 0.1$\% each. Considering that the finesse ${\cal F}$ is
\[
{\cal F} = \frac{\pi}{1-R} = \frac{\pi}{T+P},
\]
where $R+T+P = 1$, and assuming that the transmission of the mirrors $T$ are such that $T\ll P = 4\times 10^{-3}$ (four passages through the waveplates), the absorption of the wave-plates limits the finesse to
\begin{equation}
{\cal F}_{\rm max} \simeq  \frac{\pi}{P} \simeq 800.
\nonumber
\end{equation}

In this case the predicted QED ellipticity signal would be
\begin{equation}
\psi^{\rm (WavePlates)} = \left(\frac{2{\cal F_{\rm max}}}{\pi}\right) \psi^{\mbox{\scriptsize(HERA-X)}} \approx 2.5\times10^{-11}.
\nonumber
\end{equation}
Assuming for HERA-X the best birefringence sensitivity as shown in Figure \ref{noise}, this would mean a value of $S_{\Delta n}^{\rm (100~Hz)}\simeq 2.5\times10^{-20}\;1/\sqrt{\rm Hz}$ @ 100~Hz (with $\nu_{\rm L} = 25$~Hz). %For comparison, at present the PVLAS polarimeter has, at 10~Hz, a sensitivity in birefringence $S_{\Delta n}^{\rm (PVLAS)} \approx 10^{-19}\;1/\sqrt{\rm Hz}$, with a finesse of ${\cal F}^{\rm (PVLAS)} = 7\times 10^5$ and a cavity length $L^{\rm (PVLAS)} = 3.3$~m.
and a corresponding sensitivity in ellipticity $S_\psi\approx7.5\times10^{-9}\;1/\sqrt{\rm Hz}$. The corresponding integration time to reach $S/N = 1$ would therefore be
\begin{equation}
T = \left(\frac{S_{\psi}}{\psi^{\rm (WavePlates)}}\right)^{2} \lesssim 10^{5}\;{\rm s}.
\end{equation}
Such a sensitivity remains to be demonstrated in the exceptional conditions of the proposed HERA-X experiment, but with such a low finesse, near shot-noise ellipticity sensitivities have been demonstrated. Furthermore very long Fabry-Perot cavities have been shown to be stable at frequencies of a few tens of hertz by LIGO and VIRGO reaching shot-noise performances \cite{Abadie2012}. The numbers seem to be within reach and we believe that this scheme could be a viable solution when using high field static magnetic fields generated by superconducting magnets.

\section{Conclusion}

In this note we have proposed a new scheme for a sensitive polarimeter dedicated to measuring vacuum magnetic birefringence based on a Fabry-Perot cavity which would allow the use of static magnetic fields generated by superconducting magnets. The modulation of the birefringence, necessary to reach high sensitivities, is performed by two co-rotating half-wave-plates \emph{inside} the cavity, thus satisfying two conditions: rotating polarisation of the light inside the magnetic field; fixed polarisation direction on the Fabry-Perot mirrors. Furthermore the polarisation direction on the two mirrors can be controlled independently.


\begin{thebibliography}{00}

\bibitem{EHW}
H. Euler and B. Kockel, Naturwiss. {\bf23}, 246 (1935); H. Euler, Ann. Phys. (Leipzig) {\bf26}, 398 (1936); W. Heisenberg and H. Euler, Z. Phys. {\bf98}, 714 (1936); V.S. Weisskopf, K. Dan. Vidensk. Selsk., Mat. Fys. Medd. {\bf14}, 6 (1936); see also: R. Karplus and M. Neuman, Phys. Rev. {\bf80}, 380 (1950); J. Schwinger, Phys. Rev. {\bf82}, 664 (1951).
\bibitem{Delta_n_QED}
T. Erber, Nature {\bf4770}, 25 (1961); R. Baier and P. Breitenlohner, Acta Phys. Austriaca {\bf25}, 212 (1967); Nuovo Cimento {\bf47}, 117 (1967); Z. Bialynicka-Birula and I. Bialynicki-Birula, Phys. Rev. D {\bf2}, 2341 (1970); S. Adler, Ann. Phys. (NY) {\bf67}, 599 (1971).
%\bibitem{birif}
%R. Baier and P. Breitenlohner, Acta Phys. Austriaca {\bf25}, 212 (1967).\\
%R. Baier and P. Breitenlohner, Nuovo Cimento {\bf47}, 261 (1967).\\
%S.L. Adler, Ann. Phys. {\bf67} (1971) 559.\\
%Z. Bialynicka-Birula and I. Bialynicki-Birula, Phys. Rev. D {\bf2}, 2341 (1970).
\bibitem{DellaValle2015}
F. Della Valle {\em et al.} (PVLAS collaboration), Eur. Phys. J. C, in press, arXiv:1510.08052v1.
\bibitem{Q&A2010}
H.-H. Mei {\em et al.} (Q \& A collaboration), Mod. Phys. Lett. A {\bf25}, 983 (2010).
\bibitem{BMV2014}
A. Cad\`ene {\em et al.} (BMV collaboration), Eur. Phys. J. D {\bf68}, 10 (2014).
\bibitem{OSQAR2005}
P. Pugnat {\em et al.} (OSQAR collaboration), Czech J. Phys {\bf55}, A389 (2005).
\bibitem{QED2015}
{\em QED vacuum birefringence workshop}, DESY, Hamburg (D), 1-3 November 2015, https://indico.desy.de/conferenceDisplay.py?confId=12654
\bibitem{HERA-X}
J. H. P\"old, ``VMB measurement @ ALPS II. Considerations for Integrating QED optics in the ALPS II setup", presentation at the {\em QED vacuum birefringence workshop}, DESY, Hamburg (D), 1-3 November 2015.
\bibitem{ALPS2c}
R. B\"ahre {\em et al.} (ALPS collaboration), J. Instrum. {\bf 8}, T09001 (2013).
\bibitem{Trines2015}
D. Trines, ``Dipole magnet considerations for VMB. Some information on the HERA dipoles
for a magnetic birefringence measurement with the ALPS II setup ", presentation at the {\em QED vacuum birefringence workshop}, DESY, Hamburg (D), 1-3 November 2015.
\bibitem{Cameron1993}
R. Cameron {\em et al.} (BFRT collaboration), Phys. Rev. D {\bf47}, 3707 (1993).
\bibitem{Bregant2008}
E. Zavattini {\em et al.} (PVLAS collaboration), Phys. Rev. D {\bf77}, 032006 (2008); M. Bregant {\em et al.} (PVLAS collaboration), Phys. Rev. D {\bf78}, 032006 (2008).
\bibitem{DellaValle2013}
F. Della Valle {\em et al.} (PVLAS collaboration), New J. Phys. {\bf15}, 053026 (2013).
\bibitem{Zavattini2006}
G. Zavattini {\em et al.}, Appl. Phys. B {\bf83}, 571 (2006).
\bibitem{Abadie2012}
J. Abadie {\em et al.} (LIGO and VIRGO collaborations), Phys. Rev. D {\bf85}, 122007 (2012).
\end{thebibliography}
\end{document}